\documentclass[10pt,letterpaper]{article}
\usepackage[top=0.85in,left=1.6in,footskip=0.75in]{geometry}

\usepackage{amsmath,amssymb}

\usepackage{changepage}

\usepackage[utf8x]{inputenc}

\usepackage{textcomp,marvosym}

\usepackage{cite}

\usepackage{nameref,hyperref}

\usepackage[right]{lineno}

\usepackage{microtype}
\DisableLigatures[f]{encoding = *, family = * }

\usepackage[table]{xcolor}

\usepackage{array}

\usepackage{soul}

\usepackage{caption,tabularx,booktabs}
\usepackage{pdfpages}

\newcolumntype{+}{!{\vrule width 2pt}}

\newlength\savedwidth

\usepackage[aboveskip=1pt,labelfont=bf,font=small,labelsep=period,singlelinecheck=off]{caption}

\makeatletter
\renewcommand{\@biblabel}[1]{\quad#1.}
\makeatother

\usepackage{lastpage,fancyhdr,graphicx}
\usepackage{epstopdf}
\usepackage{cite}
\usepackage{amsmath,amssymb,amsfonts}
\usepackage{algorithmic}
\usepackage{dblfloatfix}
\usepackage{xcolor} %per colorare porzioni di testo
\usepackage{subcaption}	% subfigures
\usepackage{multicol}
\usepackage{graphicx}	
\usepackage{mwe}
\usepackage{textcomp}
 \usepackage{steinmetz}

\newcommand{\mypxxit}[2]{P_{{#1}}[#2]}

\newcommand{\mypxyitmk}{P_{jk}[f]}

\newcommand{\myCitmk}{C_{jk}[f]}
\newcommand{\myImCitmk}{IC_{jk}[f]}

\begin{document}
%\pagenumbering{arabic}
\vspace*{0.2in}
\begin{flushleft}

{\Large
\textbf\newline{\textbf{Phase/amplitude synchronization of brain signals during motor imagery BCI tasks}}
}

%Phase\amplitude synchronization of brain signals during motor imagery tasks

%A two-fold EEG synchronization mechanism underlies motor imagery-based BCI tasks

%Synchronization of brain signal amplitudes and phases during motor imagery-based BCI tasks

% Complementary EEG synchronization mechanisms improve motor imagery-based BCI tasks

%\title{EEG amplitude and phase synchronization characterize motor imagery-based BCI tasks}

%%Assessing functional brain connectivity during motor imagery

%Both EEG amplitude and phase synchronization characterize motor imagery-based BCI tasks.

%Amplitude and phase synchronization of EEG signals characterize motor imagery-based BCI tasks.

%EEG amplitude and phase synchronization discriminate motor imagery-based BCI tasks.

%EEG amplitude and phase synchronization improve classification of motor imagery-based BCI states.

% Complementary EEG network mechanisms of motor imagery-based BCI tasks. 

%EEG amplitude and phase synchronization in motor imagery-based BCI tasks.

% %Amplitude and phase synchronization of EEG signals characterize complementary features of motor imagery-based BCI tasks.

%\date{}

%\maketitle{}

%...connectivity, network, node degree, 
\bigskip

Tiziana Cattai\textsuperscript{1,2,3},
Stefania Colonnese\textsuperscript{3},
Marie-Constance Corsi\textsuperscript{1,2},
Danielle S. Bassett\textsuperscript{4,5,6,7,8}
Gaetano Scarano\textsuperscript{3}
Fabrizio De Vico Fallani\textsuperscript{1,2,*}
\\
\bigskip
\textbf{1} Inria Paris, Aramis Project Team, Paris, France
\\
\textbf{2} Institut du Cerveau et de la Moelle epiniere, ICM, Inserm U 1127, CNRS UMR 7225, Sorbonne Universite, Paris, France
\\
\textbf{3} Dept. of Information Engineering, Electronics and Telecommunication, Sapienza University of Rome, Italy 
\\
\textbf{4} Department of Bioengineering, University of Pennsylvania, Philadelphia, PA, 19104, USA
\\
\textbf{5} Department of Electrical and Systems Engineering, University of Pennsylvania, Philadelphia, PA, 19104, USA
\\
\textbf{6} Department of Physics \& Astronomy, University of Pennsylvania, Philadelphia, PA, 19104, USA
\\
\textbf{7} Department of Neurology, Hospital of the University of Pennsylvania, Philadelphia, PA, 19104, USA
\\
\textbf{8} Department of Psychiatry, Hospital of the University of Pennsylvania, Philadelphia, PA 19104, USA
\\
\bigskip

Corresponding author: fabrizio.devicofallani@gmail.com

\end{flushleft}
\bigskip

\section*{Abstract}

\begin{small} 
%Looking for new features is crucial to improve performance in brain-computer interfaces (BCIs).
%The selection of the features that characterize brain functioning is a crucial step for the definition of brain-computer interfaces (BCIs).
The extraction of brain functioning features is a crucial step in the definition of brain-computer interfaces (BCIs).
In the last decade, functional connectivity (FC) estimators have been increasingly explored based on their ability to capture synchronization between multivariate brain signals. 
%based features have shown the potential to discriminate different mental states. 
However, the underlying neurophysiological mechanisms and the extent to which they can improve performance in BCI-related tasks, is still poorly understood.
%Methods to expan
To address this gap in knowledge, we considered a group of $20$ healthy subjects during an EEG-based hand motor imagery (MI) task. We studied two well-established FC estimators, i.e. spectral- and imaginary-coherence, and investigated how they were modulated by the MI task. We characterized the resulting FC networks by extracting the strength of connectivity of each EEG sensor and compared the discriminant power with respect to standard power spectrum features.
%Results
At the group level, results showed that while spectral-coherence based network features were  increasing in the controlateral motor area, those based on imaginary-coherence were decreasing.
We demonstrated that this opposite, but complementary, behavior was respectively determined by the increase in amplitude and phase synchronization between the brain signals. 
At the individual level, we proved that including these network connectivity features in the classification of MI mental states led to an overall improvement in accuracy.
%of $+9.5\%$. 
%Conclusion
Taken together, our results provide fresh insights into the oscillatory mechanisms subserving brain network changes during MI and offer new perspectives to improve BCI performance.
\end{small}

%\linenumbers

\newpage
\section*{Introduction}

%Looking for new features is important to improve performances in brain-computer interfaces (BCIs).
Based on the classification of mental states from brain signals, brain-computer interfaces (BCIs) are increasingly explored for control and communication, as well as for the treatment of neurological disorders (e.g. stroke), particularly via the ability of subjects to voluntary modulate their brain activity through mental imagery.
Altough the promises, the impact of BCIs has been limited because of  their poor usability in real-life applications. BCI accuracy - as measured by correct classification of the user’s intent - is still highly variable across individuals \cite{carlson2013brain}. It is estimated that a non-negligible portion of users (around $30\%$) is not able to voluntarily modulate the brain activity and reach the accuracy level needed for minimal communication, i.e. 70\% \cite{daly2008brain,vansteensel2016fully}. This phenomenon, generally referred to as BCI-illiteracy \cite{pichiorri2015brain}, significantly limits the benefit of BCIs in most clinical scenarios \cite{lafleur2013quadcopter,soekadar2015brain}.

In the last decade many solutions have been proposed to improve BCI accuracy. On one hand, investigators have focused on the research of the best mental strategy to detect the user's intent or on the choice of the sensory feedback to convey the most relevant information to the user \cite{ahn2015performance,muller2015towards,vidaurre2011co } . On the other hand, advanced signal processing methods and sophisticated classification algorithms have been explored and developed, respectively, to improve the signal-to-noise ratio and to correctly identify the user's intent \cite{van2009comparing}. 
%Among others, linear discriminant analysis (LDA) and support vector machines (SVMs) are powerful methods that simultaneously analyze multivariate signals or features extracted from the entire sensor array \cite{van2009comparing}.
While these methods can provide considerable performance increments, they are intrinsically blind to the neural mechanisms that allowed investigators to classify the user's intent and may not have an obvious physical or physiological interpretation \cite{lotte2007review} . However, this is crucial especially in clinical settings where brain functioning can be compromised and alternative solutions must be identified.

An alternative approach would consist in looking for different - potentially more informative - features characterizing the human brain functioning. Among others, functional connectivity (FC) aims to estimate information integration between spatially distributed brain areas by measuring the temporal dependence between the regional activities \cite{fallani2014graph}. Thus, in contrast to univariate features such as frequency band power, FC appears more appropriate to capture the oscillatory network mechanisms involved in brain (re)organization during mental tasks \cite{fallani2019network}.
Recent results have demonstrated the potential of FC features in BCI, albeit the results are variable and difficult to compare because of the different FC estimators, tasks and limited number of subjects used in those studies \cite{hamedi2016electroencephalographic,
wei2007amplitude, song2006phase, brunner2005phase}. 
More importantly, the neurophysiological and mechanistic interpretation of FC features is still poorly understood in BCI-related tasks, but this is critical to assess the actual impact on accuracy and performance.

To address this question we considered two well-established FC estimators, i.e. the spectral-coherence and imaginary-coherence \cite{carter1987coherence, nolte2004identifying}. From a theoretical perspective, these estimators bring complementary information since the first measures the synchronization between the signal amplitudes while the latter is also sensitive to their phase difference  \cite{nolte2004identifying, rosenblum1996phase}.
We hypothesized that integrating these complementary features will allow a better characterization of the BCI-related mental states and that including them in the feature extraction block would serve to increase the BCI accuracy as compared to standard approaches solely based on power spectra.
To test these predictions, we considered brain FC networks derived from EEG data recorded in a group of $20$ healthy subjects performing the motor imagery (MI) of the right hand grasping. To allow a fair comparison with the results obtained with power spectrum features, we extracted for each sensor the node strength, an intuitive graph theoretic metric quantifying its overall connection intensity within the network. 
At the group level, we compared the spatial patterns extracted by statistically contrasting the feature values in the MI with respect to a baseline condition, where subjects were at rest. At the individual level, we evaluated the associated performance by means of an \textit{off-line} classification simulation. 
See \textbf{Material and methods} for more details on the experimental design and methods of analysis.

\section*{Results}
\subsection*{EEG network connectivity changes during motor imagery}
\begin{figure*}[hb]
		\centerline{\includegraphics[width=\textwidth]{{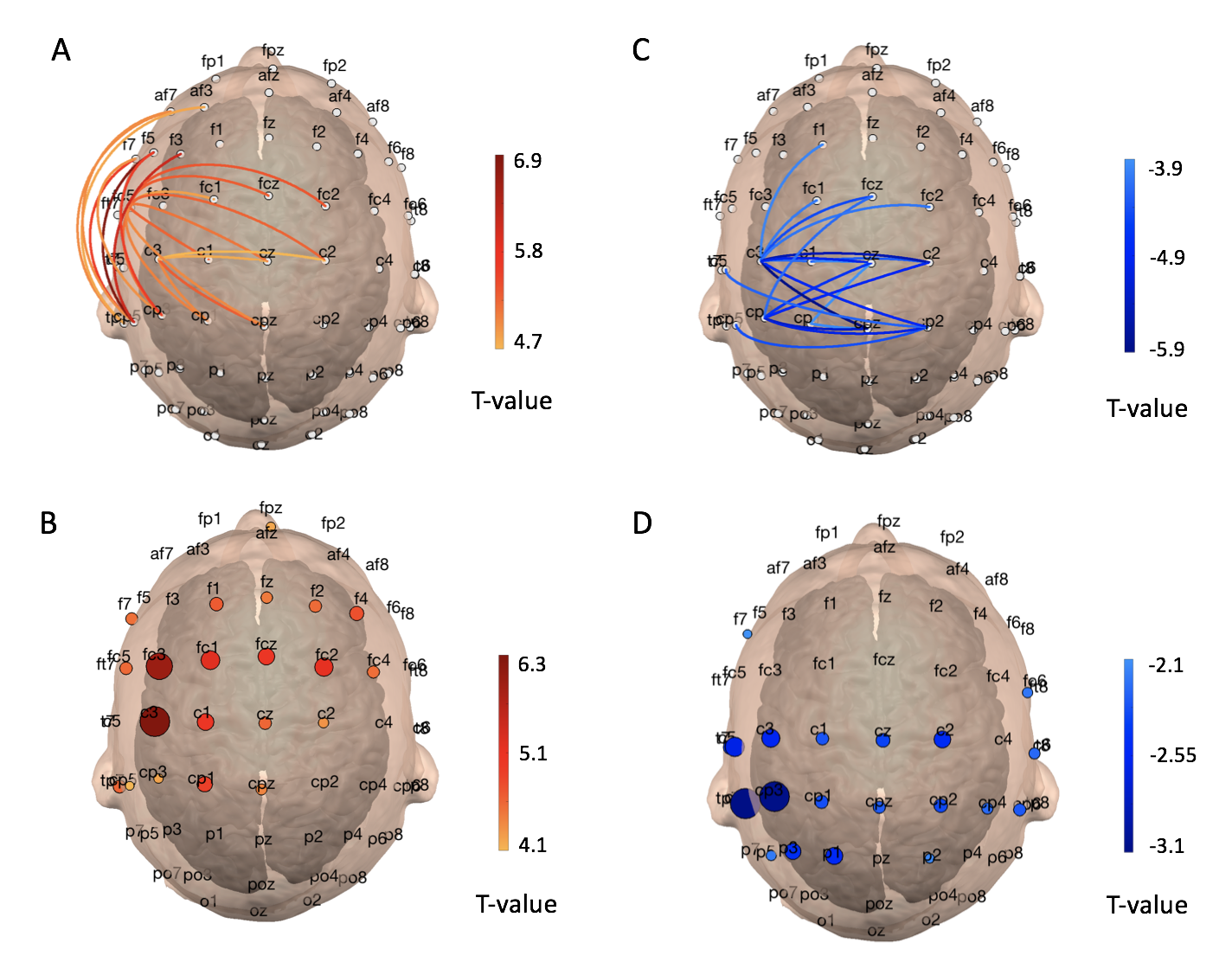}}}
		\caption{Statistical contrast maps between motor imagery and resting states in the \textit{beta} band. Permutation-based t-tests were perfromed with a statistical threshold of $0.05$ FDR-corrected for multiple comparisons. In Panel A) results for coherence, in B) for imaginary coherence, in C) for coherence-based node degree and in D) for imaginary coherence based node degree. Only the twenty most discriminant connections and nodes are represented here for the sake of simplicity.}
		\label{fig:t_test_connectivity}
\end{figure*}

To quantify the task-related changes at the group level, we considered in each subject the trial-averaged FC values and the associated node strengths $S$ (\textbf{Material and methods}). 
As expected, results showed a significant stronger involvement of the motor-related areas that are contralateral to the imagined movement ($p<0.05$, FDR-corrected). This could be appreciated both for single connection (\textbf{Fig.~\ref{fig:t_test_connectivity}}, top row) and node strength values (\textbf{Fig.~\ref{fig:t_test_connectivity}}, bottom row).

Interestingly, we found that the direction of the significant difference is opposite depending on whether we used spectral-coherence ($C$) or imaginary-coherence ($IC$) to estimate EEG networks.
We reported significant MI-related increases when we considered $C$ estimators, while we observed significant decrements when using $IC$. In terms of spatial locations these differences involved both intra-hemispheric and inter-hemispheric interactions, while the largest changes in node strength tended to concentrate around the brain areas corresponding to the EEG electrode $C3$.
The magnitude of these network changes appeared significantly higher compared to those obtained by using classical band-power features (\textbf{Fig. S1}). Furthermore, we did not report any significant correlation between the two types of features.

These findings indicated that the motor imagery of the hand grasping elicits detectable brain network changes that can be used to better characterize and discriminate MI-based BCI tasks. These changes revealed the existence of two parallel connectivity behaviors (i.e. increase for $C$ and decrease for $IC$) that primarily involve the motor areas contralateral to the movement. 

\subsection*{Modulation of amplitude and phase synchronization between brain signals}
To better understand the nature of such dichotomy, we investigated more in detail the behavior of $C$ and $IC$ estimators.
$C$ is obtained from the cross-spectrum of the two signals (\textbf{Materials and methods}) and is sensitive to the amplitude synchronization, i.e. when signals oscillate (or vary) at the same frequency. $IC$ is also sensitive to the phase synchronization capturing possible time shifts between the signals (\textbf{Material and Methods}). 

\begin{figure*}[hb]
		\centerline{\includegraphics[width=\textwidth]{{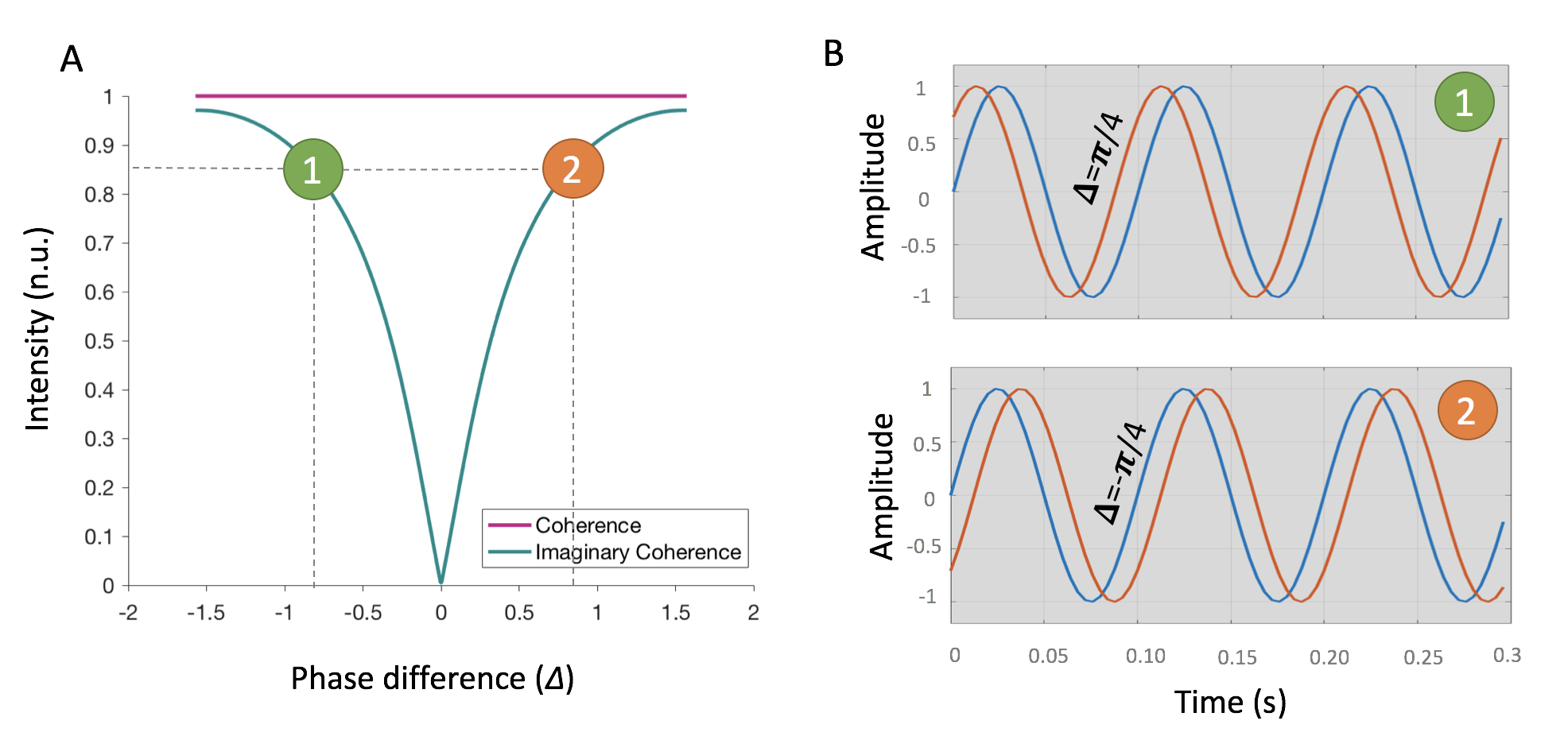}}}
		\caption{Relationships of coherence/imaginary coherence with with phase difference. In Panel A) coherence is in pink and imaginary coherence in green, showing the functional connectivity between two sines waves at $10$ Hz as function of their temporal shift. The shift, corresponding here to a phase difference, varies from $0$ to $\pi$ in steps of $\pi$ /$500$. At each shift value, the two connectivity estimators are evaluated. Panel B) shows the sine waves with different phase differences.  In panel $1$), a positive $\Delta$ of $\pi$ /4 in panel $2$), a negative $\Delta$ of –$\pi$/$4$.}
		\label{fig:lag}
\end{figure*}

To show these behaviors, we considered two perfectly equal sine waves oscillating at $10 Hz$, and we temporally shifted one with respect to the other within the [$-\pi/2$, $\pi/2$] interval. \textbf{Fig.~\ref{fig:lag}} shows that $C$ remains constant along the entire phase shift range, while $IC$ varies in a way that it tends to zero when the two signals are perfectly in phase (i.e. $\Delta=0$). In a supplemental analysis, we indeed demonstrated that the imaginary coherence between those signals can be analytically expressed as a function of their relative time delay (\textbf{Supplementary text}).
%This could be also analytically demonstrated and in the case of more complex signals it has been emphasized that the estimated imaginary coherency between two time series can be expressed as a function of the instantaneous phase difference of their analytic signals \cite{stam2007phase}  

\begin{figure*}[hb]
		\centerline{\includegraphics[width=\textwidth]{{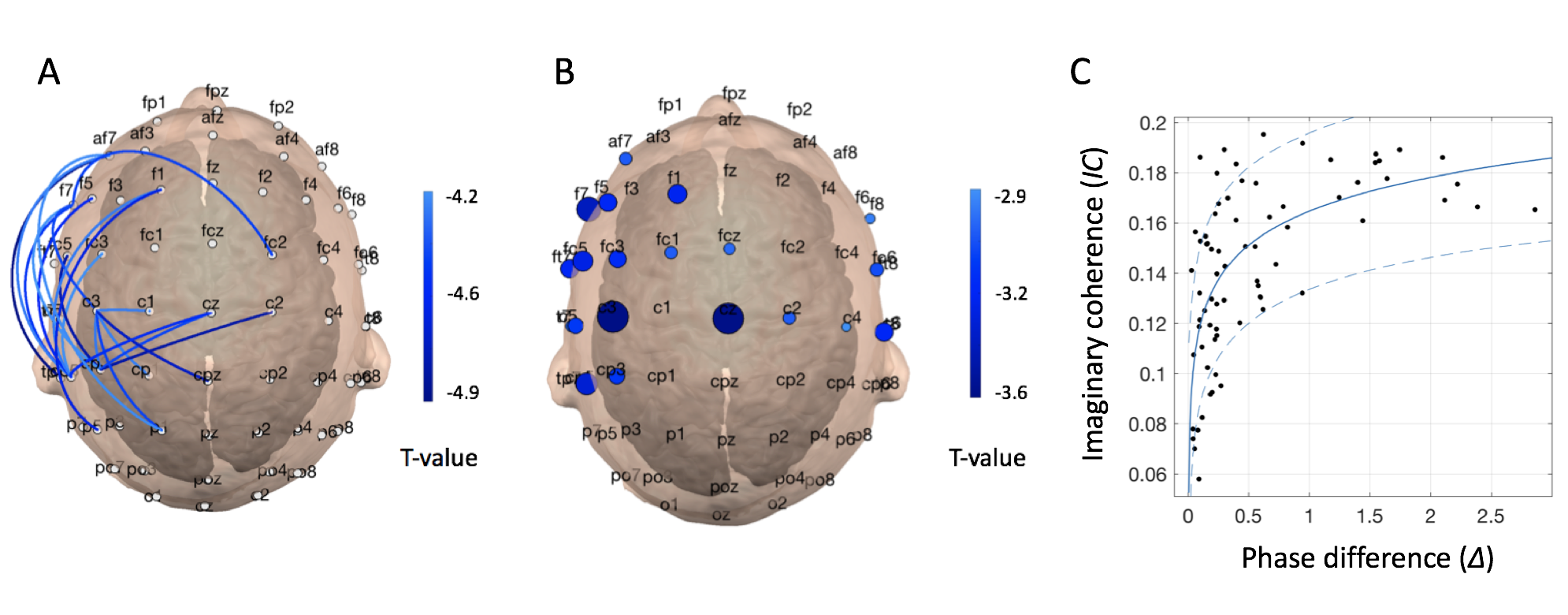}}}
		\caption{Phase difference properties and discrimination ability.  Panel A), results of permutation-based t-tests in the \textit{beta} band across all subjects are shown for brain networks reconstructed from the phase difference between EEG signals. Panel B) results of permutation-based t-tests obtained with node strength values extracted from the previous brain networks. Panel C), Spearman correlation plot between  imaginary coherence and phase difference values considering all the connections including C3 electrode for one representative subject.}
\label{fig:lag_topo}
\end{figure*}

Our experimental results showed that during MI there is a simultaneous amplitude synchronization (captured by $C$) and phase-synchronization (captured by $IC$), the latter suggesting a significant signal phase alignment (\textbf{Fig.~\ref{fig:t_test_connectivity}}). To confirm this finding, we re-estimated the task-related brain networks by computing the phase difference $\Delta$ between the EEG signals (\textbf{Material and methods}).
For both single connection and node strength statistics we reported a global significant decrease which is actually similar, in terms of magnitude and spatial arrangement, to what observed with $IC$ (\textbf{Fig.~\ref{fig:lag_topo}}A,B).

More formally, we investigated the MI task-related relationship between imaginary coherence and relative phase difference. For each subject, we considered all the pairs of nodes including the $C3$ sensor, representing the controlateral primary motor area of the hand. Results showed a moderate correlation (group-median Spearman's $R=0.38$) with lower $\Delta$ values predicting lower $IC$ ones (\textbf{Fig.~\ref{fig:lag_topo}}, \textbf{Table S1}).
 
These findings indicated that hand MI elicits a two-fold mechanism supporting more efficient information transfer - in terms of amplitude and phase synchronization - among sensorimotor brain regions.

\subsection*{Improved mental state detection in single individuals}
Finally, we tested the ability of these brain connectivity features to discriminate MI and resting states at single subject level.
To increase specificity, we considered a finer frequency resolution of $1$ Hz - from $4$ to $40$ Hz - and we restricted the feature extraction to the contralateral sensorimotor zone (\textbf{Material and  methods}).
Specifically, for each MI and rest trial we extracted three type of features: power spectrum $P$, coherence-based node strength $S^C$ and imaginary coherence-based node strength $S^{IC}$.

\begin{figure*}[hb]
		\centerline{\includegraphics[width=\textwidth]{{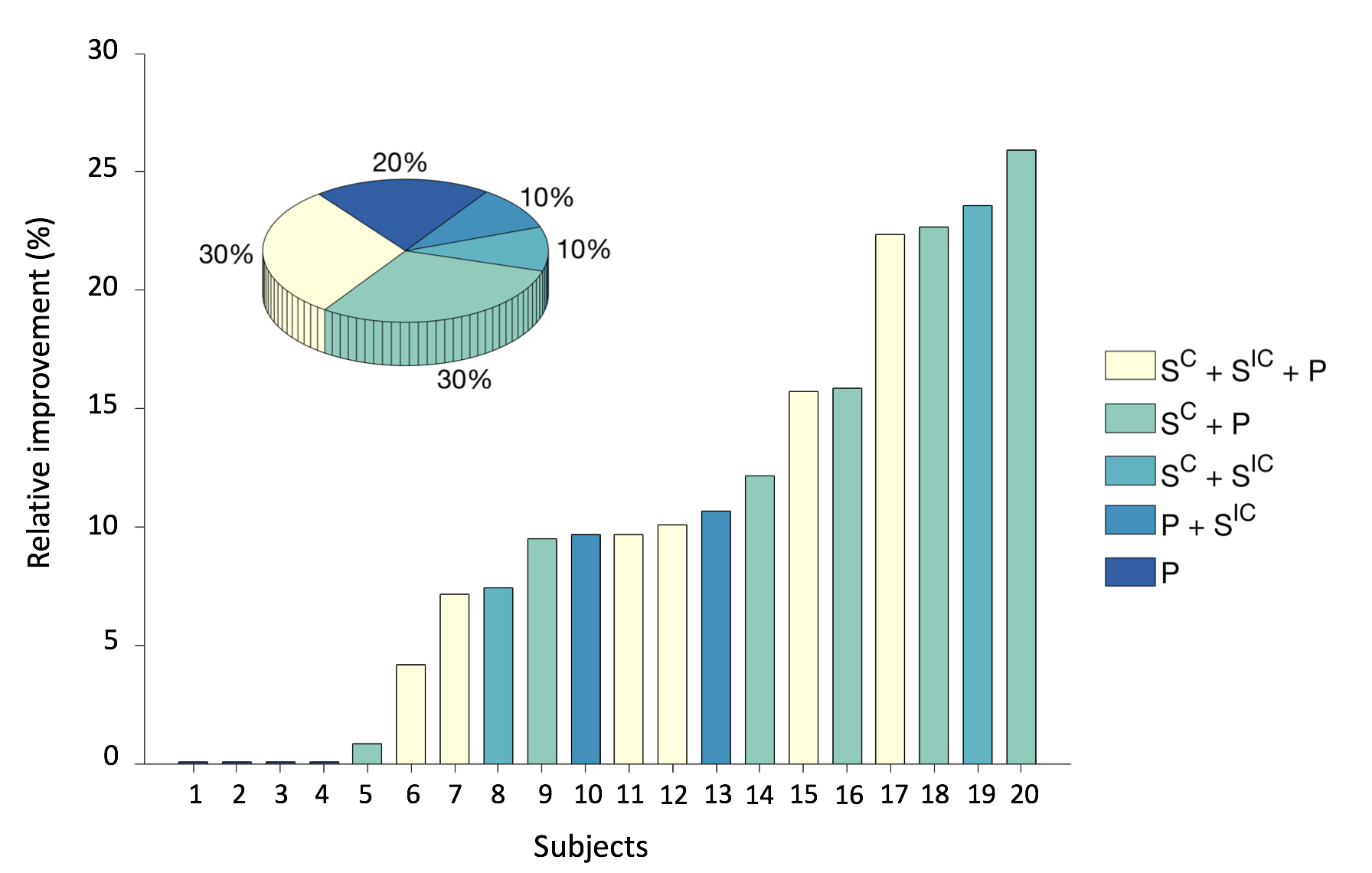}}}
		\caption{Improvement of classification performance. Bar plots show the percentage of relative increment between the best combination of features (i.e., coherence-based node strength $S^C$, imaginary coherence-based node strength $S^{IC}$, band power $P$) and band power only. The pie diagram in the inset illustrates the percentage of times that a specific combination of features has been selected across subjects.}
		\label{fig:perf}
\end{figure*}

To identify the best discriminant features, we performed a sequential forward feature selection \cite{ferri1994comparative} within a cross-validation linear discriminant analysis (LDA) (\textbf{Material and methods}). We used the overall accuracy to measure the average classification performance across validations. While in general the classification accuracy was moderate (\textbf{Table S2}), we observed that in $16/20$ subjects the inclusion of node strength features led to increase of performance in terms of relative difference with respect to $P$ features alone (\textbf{Fig.~\ref{fig:perf}}). For those subjects the  performance increment was up to $12\%$.

Notably, we observed that neither $S^C$ or $S^{IC}$ could give the best performance when considered alone. Their effect only emerged when combined together. Hence, in $30\%$ of subjects the best combination was $P$ and $S^C$, while in another $30\%$ the integration of $P$, $S^C$ and $S^{IC}$ was the best choice (\textbf{Fig.~\ref{fig:perf}}). 

\begin{figure*}[hb]
		\centerline{\includegraphics[width=\textwidth]{{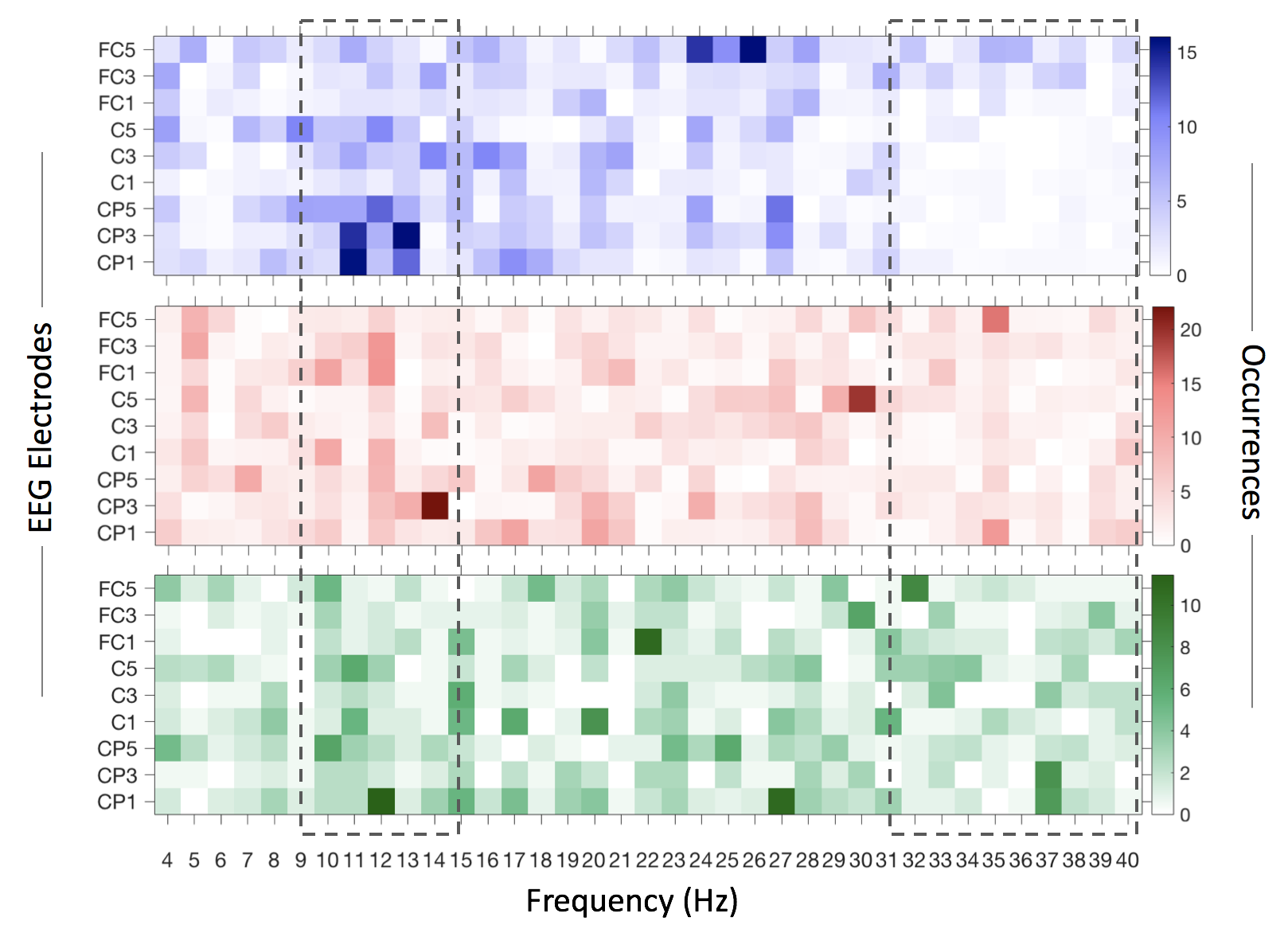}}}
		\caption{Brain features selected by the classification procedure.  The color codes for the group-averaged number of times that a specific feature - in the electrode-frequency space - has been chosen during the sequential  feature selection algorithm (\textbf{Materials and methods}). The results for $P$ features are illustrated in the top line, those for $S^C$ in the middle line and those for $S^{IC}$ in the bottom line.}
		\label{fig:occurrences}
\end{figure*}

To identify the spatial and spectral characteristics of the selected features, we showed their cumulative occurrence in a frequency-sensor plot (\textbf{Fig.~\ref{fig:occurrences}}). In general, we observed a concentration of features in the $10-14$ Hz range within the C-CP zone. For both $P$ and $S^C$, the occurrences at higher frequencies tended to fade out (\textbf{Fig.~\ref{fig:occurrences}}A,B), while the situation was more heterogenous for $S^{IC}$ features (\textbf{Fig.~\ref{fig:occurrences}}C). 

Taken together, these results indicate that brain connectivity features, capturing both amplitude and phase synchronization, can be utilized in combination with standard power spectral features to improve the detection of motor imagery mental states in healthy subjects.

\section*{Discussion}
% A) Large-scale phase-amplitude brain connectivity changes during motor tasks
Brain activity changes during motor tasks have been largely documented through invasive and noninvasive neuroimaging techniques in non-human and human primates, as well as in animal models \cite{pfurtscheller1997eeg,pfurtscheller1999event,murthy1992coherent}. 
These changes were not limited to specific brain areas, but also occur in a coherent and synchronized manner across larger spatial scales - from millimeters to centimeters -  reflecting the need for a coherent coordination of information exchanges to accomplish the task \cite{jiang2004modulation, meirovitch2015alpha, denker2018lfp,crone2006high}.
Functional connectivity methods, estimating temporal dependence between spatially remote brain areas, represent therefore a unique opportunity to study large-scale brain network changes during motor tasks from noninvasive EEG recordings.
Previous works systematically reported FC modulations in both healthy and diseased subjects \cite{de2009evaluation,lynall2010functional}. However, different FC estimators have been used in different studies and a deeper understanding of the meaning of obtained FC values was in general overlooked. As a result, a common direction and principled interpretation of the changes observed during BCI motor-related tasks is still lacking \cite{krusienski2012value}. 

To deepen this aspect, we investigated the intrinsic nature of two popular FC estimators, spectral coherence and imaginary coherence, and realized a simple motor imagery task in a group of healthy subjects. 
Our results indicate that motor imagery elicits two major parallel oscillatory phenomena in the \textit{beta} frequency band: \textit{i)} the increase of synchronization between the EEG signal amplitudes, \textit{ii)} a decrease of phase difference which means an increase of synchronization between signal phases.
%other studies with similar changes?
Both amplitude and phase synchronization increments have been respectively reported in separate studies. The former typically codes for a basic substrate of neural communication \cite{van2012neural}, while the latter occurs to further favor information binding \cite{neuper2006event}.
In our study, both the connectivity changes were region-specific and more evident in the sensorimotor areas of the brain. Notably, they only emerged at the node strength level (i.e. aggregating the information from all the nodal connections) and were not correlated with other regional measures, such as standard power spectral densities (\textbf{Fig. S1}).
While the observed network mechanisms bring new complementary information that can be used to better characterize mental states during motor tasks, more research is needed to elucidate whether these changes only reflect direct motor-related demands or, also include indirect effects due to mirror-neuron activity as well as attentional efforts associated with the task complexity \cite{murthy1992coherent}.

%B) the role of connectivity changes in BCI
The ability to discriminate different mental states from noninvasive neuroimaging recordings has concrete implications in our daily-life, from the early detection of brain diseases to the development of effective brain-computer interface applications \cite{muller2008machine}.
In the BCI context, much of the efforts has focused on the improvement of the classification algorithms, such as the recent advances in Riemannin geometry-based approaches \cite{barachant2010riemannian,gaur2018multi}. 
While these methods can in some cases ameliorate the overall classification accuracy, the improvement potential is still high and, more importantly, they generally lack of intuitive physiological interpretations \cite{lotte2007review,fallani2019network}.
%\cite{pfurtscheller1999event,donchin2000mental}.
The research of alternative features, beyond the characterization of single region activities, is therefore a fertile field with the aim of pursuing performance \cite{daly2012brain, la2014human, hamedi2015motor, brunner2006online, wei2007amplitude, krusienski2012value}.
Our results suggest that including FC network measures of brain functioning in the features extraction block, led to a better classification accuracy in almost every subject.
While the performance increments significantly varied across individuals and led in general to moderate overall accuracy ($0.63$ in average), it is important to underline that the main goal of this work was not to maximize the absolute accuracy but to assess the potential of brain connectivity properties to improve relative performance in an offline scenario.
Future studies will be crucial to identify the most appropriate classification strategy to integrate such multimodal information in an effort to optimize online MI-based BCI settings.
 %Brain-Computer Interfaces are promising means in the treatment of neuro-physiological disorders and in rehabilitation \cite{pichiorri2015brain}. 

%The main problem affecting their usability is the wrong detection of the users mental intentions. The generated brain signals exhibit complex spatio-temporal dynamics, often mixed with noise, extracting the relevant information and classifying it correctly is not an easy task. 

%Our results show that functional interactions in brain activity takes place during MI. These findings suggest that low performances of BCIs  could be determined by the partial analysis of EEG signals, which can contain more information than those used with local features. 

%It is possible that the integration of local activity and other graph centrality metrics \cite{fallani2014graph} could be more discriminant in MI task. 

%poor scores
%Concerning BCI scores, the accuracy was extremely variable across individuals, and the averaged performance was relatively low as compared to the state-of-the-art \cite{vidaurre2010towards}. Possible causes are the limited number of trials, the choice of a simple classification algorithm \cite{lotte2007review} and the high inter-subjects variability \cite{ince2007extraction}. 

%Other causes can be the wrong design and training strategy as well as psychological and cognitive factors \cite{jeunet}.

%maybe to intergate with previous one
% Understaind the reason behind better perfroance/ crucial to improve detection of mental sats/ comparison with classificaiton algorimths 

%Methodological considerations
%stationarity
Spectral coherence and imaginary coherence are FC estimators that assume the stationarity of time series within the period of interest \cite{nolte2004identifying}. In our study, we considered time windows $5$ s, which could be too long for respecting this hypothesis \cite{kaplan2005nonstationary}. We assessed the reliability of our results by computing the augmented Dickey-Fuller test \cite{kwiatkowski1992testing} and verifying that $96\%$ of all the signals were indeed stationary. More in general, for real-time BCI applications the use of shorter time windows and FC estimators that do not need stationarity assumptions (e.g. wavelets \cite{le2001comparison}, tracking algorithms \cite{ozdemir2017recursive}), would naturally allow to circumvent this issue.

%source domain
Our analysis has focused on the EEG sensor space. Coherence-based FC estimators could be affected by volume conduction distortions introducing spurious signal interactions \cite{nolte2004identifying,pascual2007coherence}. While source-reconstruction techniques could be used to attenuate such bias \cite{jatoi2014survey}, we decided to work on the sensor space for two main reasons. First, we did not have access to the individual magnetic resonance images (MRIs) necessary to have a detailed and realistic model of the head and its compartements \cite{michel2004eeg, edelman2015eeg,baillet2001electromagnetic}. Second, FC estimators can be really sensitive to signal transformations and results could be strongly dependent to the selected reconstruction algorithm \cite{mahjoory2017consistency}. A detailed analysis on the effects of source-reconstruction was, however, beyond the scope of our study. Further research is expected to better investigate the stability of our results when working at source space level.

%fusion
When combining different types of brain characteristics (i.e. power, node strengths), we performed a fusion at the feature space level \cite{ruta2000overview}. This means that the feature vectors might be of different lengths and that the performance comparison could be biased. We verified that the lengths of the selected feature vectors was in average similar, i.e. from $1$ to $2$ elements per modality. Another possibility would be to perform the fusion at the classifier level, by combining the posterior probabilities of each separate classification \cite{corsi2019integrating}. This approach will however force the research of significant features in each modality despite their absolute discriminant power. To allow a fair comparison with band-power features, we preferred not to use this approach and let the classifier identify the best absolute combination of features.

\section*{Conclusions}

Consistent with our hypothesis, we demonstrated the contribution of brain network connectivity features in detecting mental states during typical MI-based BCI tasks.
More importantly, we have discovered that hand MI is characterized by a dual connectivity phenomenon, consisting in a simultaneous  amplitude and phase synchronization of large-scale brain activity.
Taken together, our results provide fresh insights into the network mechanisms subserving brain functional changes during MI, and offer new perspectives to improve BCI performance.

\section*{Material and methods}
\subsection*{Experimental protocol and preprocessing }

Twenty healthy subjects (aged $27.60$ ± $4.01$ years, $8$ women), all right-handed, were included in the study. All subjects were recruited within the framework of a BCI training protocol and they did not present with any medical or psychological disorder. The study was approved by the ethical committee CPP-IDF-VI of Paris and each subject signed an informed consent. All participants received financial compensation for their participation.
 
The BCI experiment consisted in a standard 1D, two-target box task \cite{wolpaw2003wadsworth}. The subject was in front of a screen with a distance of $90$ cm. When the target was up, the subject was instructed to imagine moving his/her the right hand (i.e. grasping); when the target was down, the subject had to remain at rest.
EEG data were recorded with a 74-channel system, with Ag/AgCl sensors (Easycap, Germany) in a 10-10 standard configuration. 
The reference for the EEG signals were mastoid signals and the ground electrode was on the left scalpula.  Data were recorded in a shielded room. Impedances were lower than $20$ kOhms, the sampling frequency was $1$ kHz, downsampled to $250$ Hz.
For each subject we collected $64$ trials of motor imagery and $64$ trials of resting state, each of them lasted $5$s \cite{corsi2019integrating}. 

As a pre-processing we performed on the entire dataset an independent component  analysis (ICA) to eliminate ocular and cardiac artifacts on the EEG signals, via the Infomax algorithm \cite{bell1995information} available in the  Fieldtrip toolbox \cite{oostenveld2011fieldtrip}. The ICA  was operated by the visual inspection of both time signals and their associated topographies. We removed no more than two independent components.

\subsection*{Functional connectivity and brain network features }
%We collected EEG signals in motor imagery experiments where there are two mental tasks of motor imagery (MI) and resting state (RS). We recorded EEG signals in T1 trials for MI and T2 trials for RS from N channels; in each of them we collected T signal samples.

We considered two well-established functional connectivity estimators, i.e. spectral coherence ($C$) \cite{carter1987coherence} and imaginary coherence ($IC$) \cite{nolte2004identifying}. Given two EEG time series $x_j$ and $x_k$ in a time interval T, the computation of $C_{jk}$ and $IC_{jk}$ at the frequency $f$ can be respectively obtained as:

\begin{equation}
    \myCitmk=\dfrac{\left|\mypxyitmk\right|}{\left(\mypxxit{j}{f}\cdot\mypxxit{k}{f}\right)^{1/2}} 
    \label{eqn:coherence}
\end{equation}

\begin{equation}
 \myImCitmk=\dfrac{\left|{\Im\left(\mypxyitmk\right)}\right|}{\left(\mypxxit{j}{f}\cdot\mypxxit{k}{f}\right)^{1/2}}
 \label{eqn:imcoherence}
\end{equation}
\

where $P_j[f]$ contains the samples of the power spectral density $P_{jj}(e^{i\omega})$ estimated on T-length windows, i.e. ${P_{j}[f]}=\left.P_{jj}[f](e^{i\omega})\right |_{\omega=\omega_f}$, with the angular frequency $\omega_f=2\pi f/T$; and \(P_{jk}[f]\) are samples of the cross-spectrum \(P_{jk}(e^{i\omega})\)  between \(x_{j}\) and \(x_{k}\). 

These quantities are evaluated by means of with Welch's method with Hanning time windows of $T=1 s$ and an overlap of $50\%$ \cite{welch1967use}.
While $C$ has an intuitite interpretation the advantage of capturing linear correlations in the frequency domain, $IC$, by neglecting zero-lag contributions, is more robust to spurious connectivity due to volume condition \cite{nolte2004identifying}. 
For this reason, coherence is more sensitive to short-range interactions while imaginary coherence is weights more long-distance connections \cite{cattai2018characterization}.

%Another important characteristic is that coherence depends more on the cross-correlation between the two signals with respect to their difference in phase. In contrast, imaginary coherence reflects both the cross-correlation and the shifting between the two signals, weighing more their phase difference. 

To directly quantify the phase relationship between two EEG signals at the frequency $f$, we computed their phase difference $\Delta$:
%\begin{equation}
    %\left.\overline{PD}^{(i)}_{m_1\;m_2}\right|_{S_i}=\left\{\dfrac{1}{T}\sum_{t=0}^{T-1} \dfrac{1}{||B||}\sum_{k\in B}{\left|\phase{\mypxxit{m_1}{k}}-\phase{\mypxxit{m_2}{k}}\right|}\right\}
%\end{equation}

\begin{equation}
    %\left.\overline{PD}^{(i,t)}_{m_1\;m_2}\right| _{S_i}={\left|{\phi^{(i,t)}_{m_1}[k]}-{\phi^{(i,t)}_{m_2}[k]}\right|}
    {\Delta}_{jk}[f]={\left|{\phi_{j}[f]}-{\phi_{k}[f]}\right|}
    \label{eqn:delta}
\end{equation}
where ${\phi_{j}[f]}$ , ${\phi_{k}[f]}$ are the phase terms of the discrete Fourier transforms (DFTs) of \(x_{j}\) and \(x_{k}\) on T-samples windows. 

By computing $C$, $IC$ and $\Delta$ for each pair of EEG channel, we obtained symmetric $N \times N$ matrices where $N=74$ is the number of EEG channels.
These matrices correspond to fully connected and weighted networks of $N$ nodes or units and can be studied via graph theoretic tools \cite{fallani2014graph}.
Here, we focused on a simple local centrality measure, i.e. the node strength $S$, which is given by the sum of the weights of all links coming into each node. This metric describes in an intuitive way how much one brain region, or EEG channel, is connected to all the others in a certain frequency $f$.
Hence, we for each node $j$ we extracted its strength according to the different ways we constructed the network:

\begin{equation}
    S_j^C[f]=\sum_{k=1}^{N} C_{jk}[f],
\end{equation}
\begin{equation}
  S_j^{IC}[f]=\sum_{k=1}^{N} IC_{jk}[f],
\end{equation}
\begin{equation}
   S_j^{\Delta}[f]=\sum_{k=1}^{N} \Delta_{jk}[f]
\end{equation}
  
\subsection*{Statistical Analysis and Classification}
At group level, we averaged for each subject the corresponding connectivity matrices across trials and within predefined frequency bands, namely: $theta=4-7$Hz, $alpha=8-13$Hz, $beta=14-29$Hz and $gamma=30-40$Hz. Node strengths were extracted from each of these resulting networks.
Then, we statistically compared connectivity and node strength values between MI and Rest conditions by performing permutation-based paired t-tests. More specifically, for each condition we considered the distributions of values obtained from the entire population of $20$ subjects.
We set a statistical threshold of $0.05$ and we corrected multiple comparisons with a false discovery rate (FDR)\cite{benjamini1995controlling}.

%classification
At individual level, we kept the original information and we did not average the results across trials or within frequency ranges. 
We let the classification procedure to optimally select the best discriminant features for MI and Rest conditions. We only imposed some constraints to limit the research complexity. First, we considered frequency bins from 4 to 40 Hz, due to prior reports supporting their involvement in similar motor tasks \cite{neuper2001event}. 
Second, we limited the research among 9 electrodes (FC5, FC3, FC1, C5, C3, C1, CP5, CP3, CP1) spatially covering the sensorimotor area contralateral to the imagined hand movement \cite{pfurtscheller2000spatiotemporal}. 

With the aim of comparing the contribution of the three different type of features to the overall classification we considered all their possible combinations, i.e. seven in total.
In every case, we performed a 100 repeated ten-fold cross-validation test with linear discriminant analysis (LDA) \cite{lotte2007review}. 
Notably, we performed a sequential feature selection on the training folds \cite{ferri1994comparative}. To do so, features in the training folds were first sorted in a descending order according to their t-values.
This procedure allowed to select the best features predicting the data in the test fold by sequentially adding features until there is no improvement in prediction. 

\section*{Acknowledgements}
Authors would like to acknowledge Mario Chavez for useful discussion and suggestions. The research leading to these results has received
funding from the program “Investissements d’avenir” ANR-10-IAIHU-06 (Agence Nationale de la Recherche-10-IA Institut Hospitalo-Universitaire-6). FD acknowledges
support from the “Agence Nationale de la Recherche” through contract number ANR15-NEUC-0006-02. The content is solely the
responsibility of the authors and does not necessarily represent the official views of any
of the funding agencies.
\nolinenumbers

\par\null
%\selectlanguage{english}
%\nolinenumbers
\bibliography{bib_bcicon}

\begin{thebibliography}{10}

\bibitem{carlson2013brain}
Carlson T, Millan JdR.
\newblock Brain-controlled wheelchairs: a robotic architecture.
\newblock IEEE Robotics \& Automation Magazine. 2013;20(1):65--73.

\bibitem{daly2008brain}
Daly JJ, Wolpaw JR.
\newblock Brain--computer interfaces in neurological rehabilitation.
\newblock The Lancet Neurology. 2008;7(11):1032--1043.

\bibitem{vansteensel2016fully}
Vansteensel MJ, Pels EG, Bleichner MG, Branco MP, Denison T, Freudenburg ZV,
  et~al.
\newblock Fully implanted brain--computer interface in a locked-in patient with
  ALS.
\newblock New England Journal of Medicine. 2016;375(21):2060--2066.

\bibitem{pichiorri2015brain}
Pichiorri F, Morone G, Petti M, Toppi J, Pisotta I, Molinari M, et~al.
\newblock Brain--computer interface boosts motor imagery practice during stroke
  recovery.
\newblock Annals of neurology. 2015;77(5):851--865.

\bibitem{lafleur2013quadcopter}
LaFleur K, Cassady K, Doud A, Shades K, Rogin E, He B.
\newblock Quadcopter control in three-dimensional space using a noninvasive
  motor imagery-based brain--computer interface.
\newblock Journal of neural engineering. 2013;10(4):046003.

\bibitem{soekadar2015brain}
Soekadar SR, Birbaumer N, Slutzky MW, Cohen LG.
\newblock Brain--machine interfaces in neurorehabilitation of stroke.
\newblock Neurobiology of disease. 2015;83:172--179.

\bibitem{ahn2015performance}
Ahn M, Jun SC.
\newblock Performance variation in motor imagery brain--computer interface: a
  brief review.
\newblock Journal of neuroscience methods. 2015;243:103--110.

\bibitem{muller2015towards}
M{\"u}ller-Putz G, Leeb R, Tangermann M, H{\"o}hne J, K{\"u}bler A, Cincotti F,
  et~al.
\newblock Towards noninvasive hybrid brain--computer interfaces: framework,
  practice, clinical application, and beyond.
\newblock Proceedings of the IEEE. 2015;103(6):926--943.

\bibitem{vidaurre2011co}
Vidaurre C, Sannelli C, Muller KR, Blankertz B.
\newblock Co-adaptive calibration to improve BCI efficiency.
\newblock Journal of neural engineering. 2011;8(2):025009.

\bibitem{van2009comparing}
van Delden ALE, Peper CLE, Harlaar J, Daffertshofer A, Zijp NI, Nienhuys K,
  et~al.
\newblock Comparing unilateral and bilateral upper limb training: the
  ULTRA-stroke program design.
\newblock BMC neurology. 2009;9(1):57.

\bibitem{lotte2007review}
Lotte F, Congedo M, L{\'e}cuyer A, Lamarche F, Arnaldi B.
\newblock A review of classification algorithms for EEG-based brain--computer
  interfaces.
\newblock Journal of neural engineering. 2007;4(2):R1.

\bibitem{fallani2014graph}
De~Vico~Fallani F, Richiardi J, Chavez M, Achard S.
\newblock Graph analysis of functional brain networks: practical issues in
  translational neuroscience.
\newblock Phil Trans R Soc B. 2014;369(1653):20130521.

\bibitem{fallani2019network}
De~Vico~Fallani F, Bassett DS.
\newblock Network neuroscience for optimizing brain-computer interfaces.
\newblock Physics of life reviews. 2019;.

\bibitem{hamedi2016electroencephalographic}
Hamedi M, Salleh SH, Noor AM.
\newblock Electroencephalographic motor imagery brain connectivity analysis for
  BCI: a review.
\newblock Neural computation. 2016;28(6):999--1041.

\bibitem{wei2007amplitude}
Wei Q, Wang Y, Gao X, Gao S.
\newblock Amplitude and phase coupling measures for feature extraction in an
  EEG-based brain--computer interface.
\newblock Journal of Neural Engineering. 2007;4(2):120.

\bibitem{song2006phase}
Song L, Gordon E, Gysels E.
\newblock Phase synchrony rate for the recognition of motor imagery in
  brain-computer interface.
\newblock In: Advances in neural information processing systems; 2006. p.
  1265--1272.

\bibitem{brunner2005phase}
Brunner C, Graimann B, Huggins JE, Levine SP, Pfurtscheller G.
\newblock Phase relationships between different subdural electrode recordings
  in man.
\newblock Neuroscience letters. 2005;375(2):69--74.

\bibitem{carter1987coherence}
Carter GC.
\newblock Coherence and time delay estimation.
\newblock Proceedings of the IEEE. 1987;75(2):236--255.

\bibitem{nolte2004identifying}
Nolte G, Bai O, Wheaton L, Mari Z, Vorbach S, Hallett M.
\newblock Identifying true brain interaction from EEG data using the imaginary
  part of coherency.
\newblock Clinical neurophysiology. 2004;115(10):2292--2307.

\bibitem{rosenblum1996phase}
Rosenblum MG, Pikovsky AS, Kurths J.
\newblock Phase synchronization of chaotic oscillators.
\newblock Physical review letters. 1996;76(11):1804.

\bibitem{ferri1994comparative}
Ferri F, Pudil P, Hatef M, Kittler J.
\newblock Comparative study of techniques for large-scale feature selection.
\newblock In: Machine Intelligence and Pattern Recognition. vol.~16. Elsevier;
  1994. p. 403--413.

\bibitem{pfurtscheller1997eeg}
Pfurtscheller G.
\newblock EEG event-related desynchronization (ERD) and synchronization (ERS).
\newblock Electroencephalography and Clinical Neurophysiology. 1997;1(103):26.

\bibitem{pfurtscheller1999event}
Pfurtscheller G, Da~Silva FL.
\newblock Event-related EEG/MEG synchronization and desynchronization: basic
  principles.
\newblock Clinical neurophysiology. 1999;110(11):1842--1857.

\bibitem{murthy1992coherent}
Murthy VN, Fetz EE.
\newblock Coherent 25-to 35-Hz oscillations in the sensorimotor cortex of awake
  behaving monkeys.
\newblock Proceedings of the National Academy of Sciences of the United States
  of America. 1992;89(12):5670.

\bibitem{jiang2004modulation}
Jiang T, He Y, Zang Y, Weng X.
\newblock Modulation of functional connectivity during the resting state and
  the motor task.
\newblock Human brain mapping. 2004;22(1):63--71.

\bibitem{meirovitch2015alpha}
Meirovitch Y, Harris H, Dayan E, Arieli A, Flash T.
\newblock Alpha and beta band event-related desynchronization reflects
  kinematic regularities.
\newblock Journal of Neuroscience. 2015;35(4):1627--1637.

\bibitem{denker2018lfp}
Denker M, Zehl L, Kilavik BE, Diesmann M, Brochier T, Riehle A, et~al.
\newblock LFP beta amplitude is linked to mesoscopic spatio-temporal phase
  patterns.
\newblock Scientific reports. 2018;8(1):5200.

\bibitem{crone2006high}
Crone NE, Sinai A, Korzeniewska A.
\newblock High-frequency gamma oscillations and human brain mapping with
  electrocorticography.
\newblock Progress in brain research. 2006;159:275--295.

\bibitem{de2009evaluation}
De~Vico~Fallani F, Astolfi L, Cincotti F, Mattia D, La~Rocca D, Maksuti E,
  et~al.
\newblock Evaluation of the brain network organization from EEG signals: a
  preliminary evidence in stroke patient.
\newblock The Anatomical Record: Advances in Integrative Anatomy and
  Evolutionary Biology: Advances in Integrative Anatomy and Evolutionary
  Biology. 2009;292(12):2023--2031.

\bibitem{lynall2010functional}
Lynall ME, Bassett DS, Kerwin R, McKenna PJ, Kitzbichler M, Muller U, et~al.
\newblock Functional connectivity and brain networks in schizophrenia.
\newblock Journal of Neuroscience. 2010;30(28):9477--9487.

\bibitem{krusienski2012value}
Krusienski DJ, McFarland DJ, Wolpaw JR.
\newblock Value of amplitude, phase, and coherence features for a sensorimotor
  rhythm-based brain--computer interface.
\newblock Brain research bulletin. 2012;87(1):130--134.

\bibitem{van2012neural}
van Wijk B, Beek PJ, Daffertshofer A.
\newblock Neural synchrony within the motor system: what have we learned so
  far?
\newblock Frontiers in human neuroscience. 2012;6:252.

\bibitem{neuper2006event}
Neuper C, Klimesch W.
\newblock Event-related dynamics of brain oscillations. vol. 159.
\newblock Elsevier; 2006.

\bibitem{muller2008machine}
M{\"u}ller KR, Tangermann M, Dornhege G, Krauledat M, Curio G, Blankertz B.
\newblock Machine learning for real-time single-trial EEG-analysis: from
  brain--computer interfacing to mental state monitoring.
\newblock Journal of neuroscience methods. 2008;167(1):82--90.

\bibitem{barachant2010riemannian}
Barachant A, Bonnet S, Congedo M, Jutten C.
\newblock Riemannian geometry applied to BCI classification.
\newblock In: International Conference on Latent Variable Analysis and Signal
  Separation. Springer; 2010. p. 629--636.

\bibitem{gaur2018multi}
Gaur P, Pachori RB, Wang H, Prasad G.
\newblock A multi-class EEG-based BCI classification using multivariate
  empirical mode decomposition based filtering and Riemannian geometry.
\newblock Expert Systems with Applications. 2018;95:201--211.

\bibitem{daly2012brain}
Daly I, Nasuto SJ, Warwick K.
\newblock Brain computer interface control via functional connectivity
  dynamics.
\newblock Pattern recognition. 2012;45(6):2123--2136.

\bibitem{la2014human}
La~Rocca D, Campisi P, Vegso B, Cserti P, Kozmann G, Babiloni F, et~al.
\newblock Human brain distinctiveness based on EEG spectral coherence
  connectivity.
\newblock IEEE transactions on Biomedical Engineering. 2014;61(9):2406--2412.

\bibitem{hamedi2015motor}
Hamedi M, Salleh SH, Samdin SB, Noor AM.
\newblock Motor imagery brain functional connectivity analysis via coherence.
\newblock In: 2015 IEEE International conference on signal and image processing
  applications (ICSIPA). IEEE; 2015. p. 269--273.

\bibitem{brunner2006online}
Brunner C, Scherer R, Graimann B, Supp G, Pfurtscheller G.
\newblock Online control of a brain-computer interface using phase
  synchronization.
\newblock IEEE Transactions on Biomedical Engineering. 2006;53(12):2501--2506.

\bibitem{kaplan2005nonstationary}
Kaplan AY, Fingelkurts AA, Fingelkurts AA, Borisov SV, Darkhovsky BS.
\newblock Nonstationary nature of the brain activity as revealed by EEG/MEG:
  methodological, practical and conceptual challenges.
\newblock Signal processing. 2005;85(11):2190--2212.

\bibitem{kwiatkowski1992testing}
Kwiatkowski D, Phillips PC, Schmidt P, Shin Y.
\newblock Testing the null hypothesis of stationarity against the alternative
  of a unit root: How sure are we that economic time series have a unit root?
\newblock Journal of econometrics. 1992;54(1-3):159--178.

\bibitem{le2001comparison}
Le~Van~Quyen M, Foucher J, Lachaux JP, Rodriguez E, Lutz A, Martinerie J,
  et~al.
\newblock Comparison of Hilbert transform and wavelet methods for the analysis
  of neuronal synchrony.
\newblock Journal of neuroscience methods. 2001;111(2):83--98.

\bibitem{ozdemir2017recursive}
Ozdemir A, Bernat EM, Aviyente S.
\newblock Recursive tensor subspace tracking for dynamic brain network
  analysis.
\newblock IEEE Transactions on Signal and Information Processing over Networks.
  2017;3(4):669--682.

\bibitem{pascual2007coherence}
Pascual-Marqui RD.
\newblock Coherence and phase synchronization: generalization to pairs of
  multivariate time series, and removal of zero-lag contributions.
\newblock arXiv preprint arXiv:07061776. 2007;.

\bibitem{jatoi2014survey}
Jatoi MA, Kamel N, Malik AS, Faye I, Begum T.
\newblock A survey of methods used for source localization using EEG signals.
\newblock Biomedical Signal Processing and Control. 2014;11:42--52.

\bibitem{michel2004eeg}
Michel CM, Murray MM, Lantz G, Gonzalez S, Spinelli L, de~Peralta RG.
\newblock EEG source imaging.
\newblock Clinical neurophysiology. 2004;115(10):2195--2222.

\bibitem{edelman2015eeg}
Edelman BJ, Baxter B, He B.
\newblock EEG source imaging enhances the decoding of complex right-hand motor
  imagery tasks.
\newblock IEEE Transactions on Biomedical Engineering. 2015;63(1):4--14.

\bibitem{baillet2001electromagnetic}
Baillet S, Mosher JC, Leahy RM.
\newblock Electromagnetic brain mapping.
\newblock IEEE Signal processing magazine. 2001;18(6):14--30.

\bibitem{mahjoory2017consistency}
Mahjoory K, Nikulin VV, Botrel L, Linkenkaer-Hansen K, Fato MM, Haufe S.
\newblock Consistency of EEG source localization and connectivity estimates.
\newblock Neuroimage. 2017;152:590--601.

\bibitem{ruta2000overview}
Ruta D, Gabrys B.
\newblock An overview of classifier fusion methods.
\newblock Computing and Information systems. 2000;7(1):1--10.

\bibitem{corsi2019integrating}
Corsi MC, Chavez M, Schwartz D, Hugueville L, Khambhati AN, Bassett DS, et~al.
\newblock Integrating eeg and meg signals to improve motor imagery
  classification in brain--computer interface.
\newblock International journal of neural systems. 2019;29(01):1850014.

\bibitem{wolpaw2003wadsworth}
Wolpaw JR, McFarland DJ, Vaughan TM, Schalk G.
\newblock The Wadsworth Center brain-computer interface (BCI) research and
  development program.
\newblock IEEE Transactions on Neural Systems and Rehabilitation Engineering.
  2003;11(2):1--4.

\bibitem{bell1995information}
Bell AJ, Sejnowski TJ.
\newblock An information-maximization approach to blind separation and blind
  deconvolution.
\newblock Neural computation. 1995;7(6):1129--1159.

\bibitem{oostenveld2011fieldtrip}
Oostenveld R, Fries P, Maris E, Schoffelen JM.
\newblock FieldTrip: open source software for advanced analysis of MEG, EEG,
  and invasive electrophysiological data.
\newblock Computational intelligence and neuroscience. 2011;2011:1.

\bibitem{welch1967use}
Welch P.
\newblock The use of fast Fourier transform for the estimation of power
  spectra: a method based on time averaging over short, modified periodograms.
\newblock IEEE Transactions on audio and electroacoustics. 1967;15(2):70--73.

\bibitem{cattai2018characterization}
Cattai T, Colonnese S, Corsi MC, Bassett DS, Scarano G, De~Vico~Fallani F.
\newblock Characterization of mental states through node connectivity between
  brain signals.
\newblock In: Signal Processing Conference (EUSIPCO), 2018 26th European. IEEE;
  2018. p. 1391--1395.

\bibitem{benjamini1995controlling}
Benjamini Y, Hochberg Y.
\newblock Controlling the false discovery rate: a practical and powerful
  approach to multiple testing.
\newblock Journal of the Royal statistical society: series B (Methodological).
  1995;57(1):289--300.

\bibitem{neuper2001event}
Neuper C, Pfurtscheller G.
\newblock Event-related dynamics of cortical rhythms: frequency-specific
  features and functional correlates.
\newblock International journal of psychophysiology. 2001;43(1):41--58.

\bibitem{pfurtscheller2000spatiotemporal}
Pfurtscheller G.
\newblock Spatiotemporal ERD/ERS patterns during voluntary movement and motor
  imagery.
\newblock In: Supplements to Clinical neurophysiology. vol.~53. Elsevier; 2000.
  p. 196--198.

\end{thebibliography}
%\clearpage
\begin{small}
\bibliographystyle{plos2015}
\end{small}

\newpage
\section*{Figures}

\newpage

\newpage

\newpage

\newpage

\newpage
\nolinenumbers



\section*{Supplementary material}

# Supplementary text

Let $x_j$ and $x_k$ be two signals of length $T$ such that they only differ of a time shift $t_0$. In this simple situation, their cross-spectrum in the continuous domain reads as

$$P_{jk}(e^{i\omega}) = P_j(e^{i\omega})e^{i\omega t_0} \tag{1}$$

where $\omega$ is the angular frequency of a signal [?]. In the discrete domain, this can be rewritten as

$$P_{jk}[f] = P_j[f]e^{i2\pi f t_0/T} \tag{2}$$

where the time shift $t_0$ becomes a linear phase term.

In the mathematical formulation of coherence $C$

$$C_{jk}[f] = \frac{|P_{jk}[f]|}{(P_j[f] \cdot P_k[f])^{1/2}} \tag{3}$$

the numerator is the real part of the cross-spectrum and the exponential term in Eq. 2 is canceled out. This indicates that $C$ values do not depend on the amount of time shift between the signals.

Instead, in imaginary coherence $IC$,

$$IC_{jk}[f] = \frac{|\Im(P_{jk}[f])|}{(P_j[f] \cdot P_k[f])^{1/2}} \tag{4}$$

it is trivial to show that there is a remaining term related to $t_0$ in the numerator. Indeed, by rewriting the cross-spectrum via trigonometric functions:

$$P_{jk}[f] = P_j[f](\cos(2\pi f t_0/T) + i\sin(2\pi f t_0/T)) \tag{5}$$

Hence, by taking the imaginary part one obtains

$$\Im(P_{jk}[f]) = P_j[f]\sin(2\pi f t_0/T) \tag{6}$$

This indicates that $IC$ values do depend on the relative delay between the signals in a very specific way. More in general, in it has been emphasized that the estimated imaginary coherency between two time series can be expressed as a function of the instantaneous phase difference of their analytic signals [1].

# References

1. Stam CJ, Nolte G, Daffertshofer A. Phase lag index: assessment of functional connectivity from multi channel EEG and MEG with diminished bias from common sources. Human brain mapping. 2007;28(11):1178–1193.

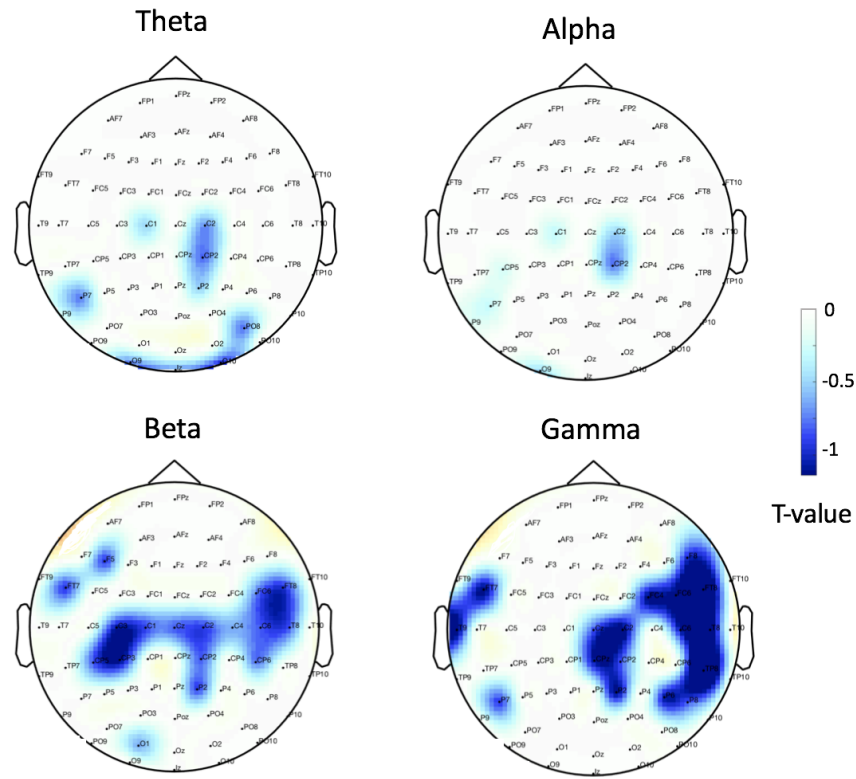

**Fig S1.** Statistical contrast maps between motor imagery and resting states obtained with band power features. Results are shown for one-tailed permutation-based t-tests. In Panel A) the obtained t-values are illustrated for *theta* band, in B) for *alpha* band, in C) for *beta* band and in *gamma* band. Despite tendencies, no significant results ($p < 0.05$, FDR-corrected) were reported.

## Subjects

| | 1 | 2 | 3 | 4 | 5 | 6 | 7 | 8 | 9 | 10 | 11 | 12 | 13 | 14 | 15 | 16 | 17 | 18 | 19 | 20 |
|---|---|---|---|---|---|---|---|---|---|---|---|---|---|---|---|---|---|---|---|---|
| Correlation ($IC$, $P$) | 0,49 | 0,19 | 0,46 | 0,54 | 0,43 | 0,35 | 0,46 | 0,14 | 0,40 | 0,28 | 0,55 | 0,22 | 0,58 | 0,46 | 0,37 | 0,23 | 0,58 | 0,15 | 0,34 | 0,20 |

**Table S1.** Table of correlation coefficients. Spearman correlation coefficient between $S^{IC}$ and $P$ for nodes including C3 for each subject.

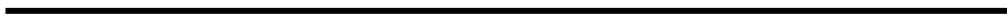

Subjects

| | 1 | 2 | 3 | 4 | 5 | 6 | 7 | 8 | 9 | 10 | 11 | 12 | 13 | 14 | 15 | 16 | 17 | 18 | 19 | 20 |
|---|---|---|---|---|---|---|---|---|---|---|---|---|---|---|---|---|---|---|---|---|
| $S^C + S^{IC} + P$ | 0,66 | **0,62** | **0,62** | 0,57 | **0,59** | 0,55 | 0,58 | **0,7** | 0,63 | 0,56 | 0,69 | 0,63 | 0,55 | **0,53** | 0,66 | **0,57** | 0,43 | 0,58 | 0,68 | 0,5 |
| $P$ | 0,59 | 0,56 | 0,53 | **0,61** | 0,55 | 0,54 | 0,58 | 0,67 | 0,58 | 0,54 | 0,63 | 0,65 | **0,59** | 0,48 | 0,62 | 0,46 | **0,54** | 0,5 | 0,56 | **0,56** |
| $S^C$ | 0,72 | 0,5 | 0,6 | 0,51 | 0,53 | 0,59 | 0,57 | 0,62 | 0,6 | 0,53 | 0,69 | 0,61 | 0,57 | 0,53 | 0,64 | 0,53 | 0,5 | 0,57 | 0,67 | 0,53 |
| $S^{IC}$ | 0,49 | 0,56 | 0,6 | 0,56 | 0,53 | 0,52 | 0,53 | 0,55 | 0,6 | 0,56 | 0,67 | 0,54 | 0,44 | 0,51 | 0,57 | 0,56 | 0,5 | 0,46 | 0,64 | 0,51 |
| $S^C + P$ | **0,74** | 0,57 | 0,6 | 0,58 | 0,53 | **0,63** | **0,59** | 0,65 | 0,56 | 0,52 | 0,65 | **0,71** | 0,58 | 0,49 | **0,69** | 0,54 | 0,43 | **0,62** | 0,67 | 0,53 |
| $S^C + S^{IC}$ | 0,65 | 0,58 | 0,6 | 0,55 | 0,57 | 0,58 | 0,58 | 0,62 | 0,62 | **0,59** | 0,68 | 0,58 | 0,52 | 0,52 | 0,56 | 0,57 | 0,45 | 0,56 | **0,7** | 0,49 |
| $P + S^{IC}$ | 0,54 | 0,58 | 0,6 | 0,5 | 0,53 | 0,54 | 0,55 | 0,69 | **0,64** | 0,53 | **0,7** | 0,56 | 0,57 | 0,52 | 0,64 | 0,56 | 0,46 | 0,5 | 0,62 | 0,53 |

**Table S2.** Table of accuracy. Average accuracy across cross-validation is reported for each subject and each combination of feature.

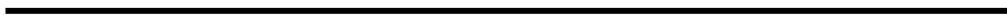



\end{document}